\documentclass[prd,reprint,superscriptaddress]{revtex4-1}

\usepackage[colorlinks=true]{hyperref}
\usepackage{amsmath}
\usepackage{graphicx}

\begin{document}

\title{Current status of direct dark matter detection experiments}
\author{Jianglai Liu}
\author{Xun Chen}
\affiliation{INPAC, Department of Physics and Astronomy, Shanghai Jiao Tong University, Shanghai Laboratory for Particle Physics and Cosmology, Shanghai 200240, China}
\author{Xiangdong Ji}
\email[Corresponding author: ]{xdji@sjtu.edu.cn}
\affiliation{INPAC, Department of Physics and Astronomy, Shanghai Jiao Tong University, Shanghai Laboratory for Particle Physics and Cosmology, Shanghai 200240, China}
\affiliation{Tsung-Dao Lee Institute, Shanghai, China}
\affiliation{Department of Physics, University of Maryland, College Park, Maryland 20742, USA}
\date{accepted 13 Jan. 2017, published 2 Mar. 2017}
\begin{abstract}
  Much like ordinary matter, dark matter might consist of elementary
  particles, and weakly interacting massive particles are one of the
  prime suspects. During the past decade, the sensitivity of
  experiments trying to directly detect them has improved by three to
  four orders of magnitude, but solid evidence for their existence is
  yet to come. We overview the recent progress in direct dark matter
  detection experiments and discuss future directions.
\end{abstract}

\maketitle

Almost a century ago, astronomers suggested the existence of a
hypothetical dark matter, invisible in the entire electromagnetic
spectrum. To date, the dominance of dark matter and its role in
driving the evolution and landscape of our universe have become the
standard paradigm in cosmology~\cite{Bertone:2004pz}. On the galactic
scale, astronomers believe that our galaxy is surrounded by an
extended and diffusive dark matter
halo~\cite{Savage:2006qr,Jungman:1995df}. Microscopically, weakly
interacting massive particles (WIMPs), with typical masses around the
electroweak symmetry breaking scale (∼100 GeV/c$^2$), are a generic
class of dark matter candidates favoured by many theories beyond the
Standard Model of particle physics~\cite{Smith:2006ym}, such as
supersymmetry (SUSY). The production and annihilation of such
particles in the early thermal universe could naturally explain the
abundance of regular matter observed today -- an appealing scenario known
as the WIMP miracle.

Another popular particle dark matter candidate is the axion, an
extremely light bosonic particle originally introduced to explain the
so-called charge-parity (CP) symmetry in the strong
interaction~\cite{Peccei:1977hh,Wilczek:1977pj}. The extremely small
coupling of axions to electromagnetic radiation and regular matter
(such as electrons and nucleons) would allow them to be detected in
ground experiments~\cite{Kim:1986ax}. However, no positive signals
have shown up so far (see ref.~\cite{Marsh:2015xka} for a recent
review). WIMPs, on the other hand, have finite weak coupling with
regular matter. They could be detected via their elastic scattering
off matter (direct detection), in high-energy collisions (collider
searches), or via their annihilation into normal particle-antiparticle
pairs (indirect detection)~\cite{Gaskins:2016cha}. In recent years,
significant experimental progress has been made in the direct
detection of WIMPs, which is the focus of this short review.

Direct detection experiments are designed to detect the nuclear recoil
in the scattering of galactic WIMPs off target nuclei. The signal rate
depends on the local density and velocity distribution of WIMPs in the
Milky Way (astrophysical inputs with non-negligible systematic
uncertainties), the WIMP mass, and the interaction cross-section of
the target nuclei~\cite{Lewin:1995rx}. Most theoretical models predict
that this cross-section is smaller than $10^{-42}$ cm$^2$ for nearly
all possible WIMP masses, yielding an extremely low signal rate, which
in turn requires an extraordinarily low background environment for
detection. To suppress the background produced by cosmic rays, all the
direct detection experiments are located in deep underground
laboratories. The residual background also includes neutrons and gamma
rays from the environment and detectors. Passive shielding and, in
some cases, active veto are required to suppress the external
background, and high-purity detector components are a must to minimize
the more dangerous internal backgrounds. The ultimate irreducible
background comes from solar and atmospheric neutrinos. The uncertainty
in neutrino-nucleus coherent scattering eventually limits the
sensitivity of the direct detection
experiments~\cite{Strigari:2009bq,Gutlein:2010tq,Ruppin:2014bra}.

\begin{figure}[htbp]
  \centering
  \includegraphics[width=0.95\linewidth]{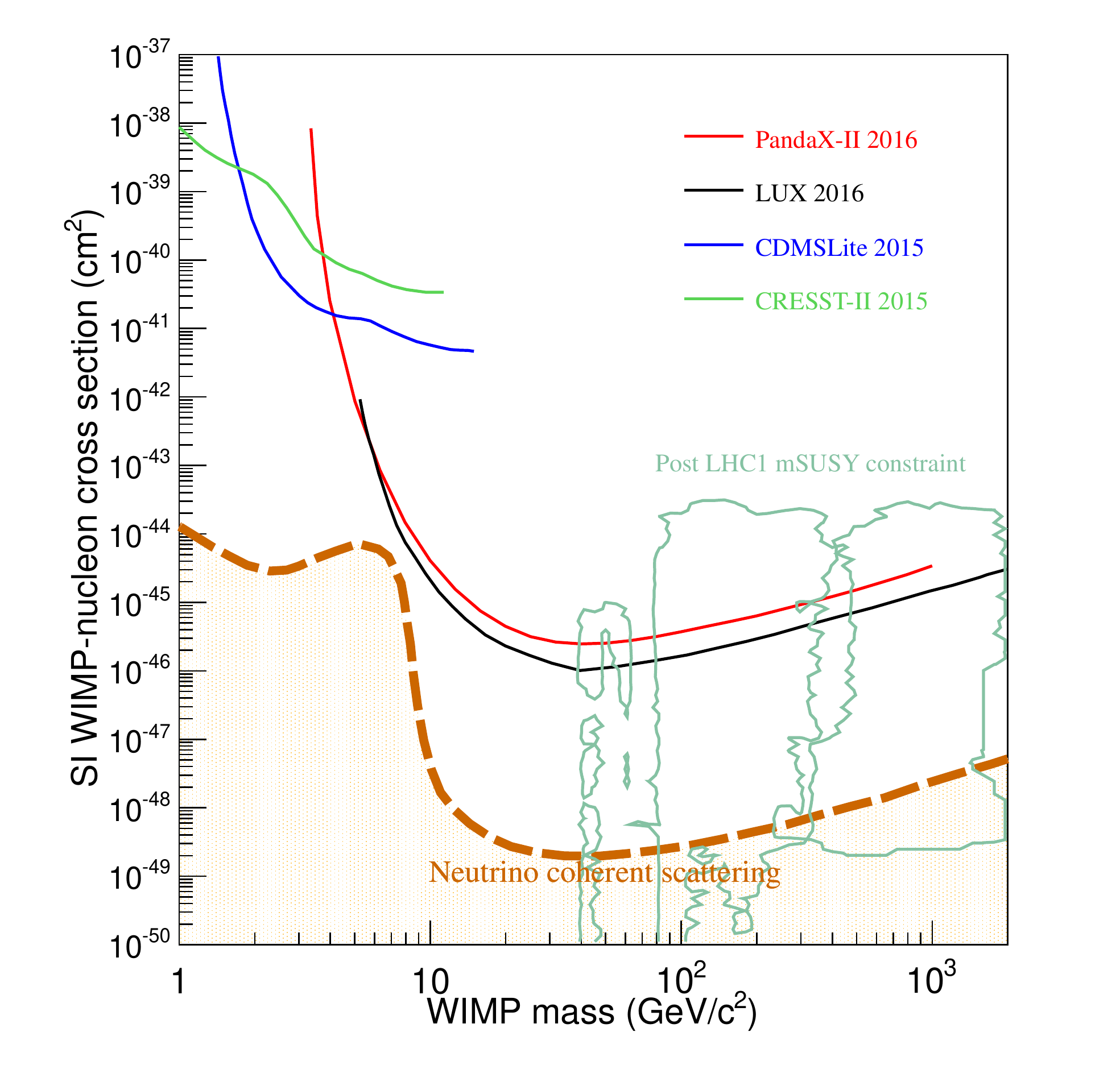}
  \caption{Upper limits on the spin-independent (SI) WIMP-nucleon
    scattering cross-section set by current leading experiments. The
    limit curves are from PandaX-II\cite{Tan:2016zwf}, LUX~\cite{Akerib:2016vxi}, SuperCDMS (CDMSLite)~\cite{Agnese:2015nto}
    and CRESST-II~\cite{Angloher:2015ewa}. The neutrino coherent scattering background curve
    data is from ref. \cite{Billard:2013qya} and the post-LHC-Run1
    minimal-SUSY model allowed contours are from
    ref.~\cite{Bagnaschi:2015eha}.}
  \label{fig:limit_si_current}
\end{figure}

For a WIMP mass between 1 and 1,000 GeV, the typical elastic recoil
energy of an atomic nucleus ranges from 1 to 100 keV (for a large
nucleus, the smaller the WIMP mass, the lower the mean recoil energy
and vice versa), which is the primary signal in direct
detection. WIMPs can in principle scatter off electrons, but the
recoil energy would be suppressed further by the electron mass and
hence would require a more sensitive detection
technology~\cite{Dedes:2009bk}. The nuclear recoil energy can be
converted into thermal motion (phonons), ionization, or scintillation
photons through the Coulomb field of the charged
nucleus~\cite{Gaitskell:2004gd}. For example, semiconductor detectors
can detect the electrons from the ionization of a heavy ion, and when
operating at very low temperatures, the same detector can be used as a
bolometer to measure the phonons produced by atomic
collisions. Experiments that simultaneously detect two types of
signals are typically more powerful in differentiating nuclear recoil
signals against electron recoil backgrounds from radioactivity. Other
features of the signals, such as the timing shape, can sometimes also
be used for background discriminations. To compare the results from
different experiments, the standard assumption is that WIMPs scatter
coherently and elastically off all nucleons in the nucleus, and the
interaction has neither nuclear spin nor a proton-neutron
dependence. To date, apart from a few controversial
claims~\cite{Bernabei:2013xsa, Aalseth:2012if, Angloher:2011uu,
  Agnese:2013rvf}, no solid WIMP signal has been observed in a direct
detection experiment. However, these placed tight constraints on
various theoretical models. Figure ~\ref{fig:limit_si_current}
summarizes the current leading direct detection limits on
spin-independent WIMP-nucleon cross-section versus the mass of the
WIMPs. Overlaid in the figure are the WIMP search sensitivity limited
by the neutrino background~\cite{Billard:2013qya}, and the
representative minimal supersymmetric model contours (2$\sigma$)
constrained by Run 1 at the Large Hadron
Collider~\cite{Bagnaschi:2015eha}. In what follows, we review some of
the progress made by a few representative experiments.

Detectors made of noble liquids are spearheading WIMP searches in the
$\sim$100 GeV/c$^2$ mass range. The so-called dual-phase XENON10
collaboration~\cite{Angle:2007uj}, have developed into the
state-of-the-art detection technology and have been pushing the
elastic spin-independent WIMP-nucleon scattering sensitivity in a wide
range of WIMP masses above 5 GeV/c$^2$
(refs~\cite{Aprile:2013doa,Akerib:2013tjd,Akerib:2015rjg,Tan:2016zwf}). Natural
xenon does not have long-lived radioactive isotopes, except for
$^{136}$Xe, whose double-beta decay has a negligible contribution to
the current generation of experiments. Liquid-xenon targets allow for
a relatively straightforward scaling-up to large monolithic
detectors. In a liquid-xenon time-projection chamber, with two arrays
of photomultiplier tubes located at its top and bottom and with a
large electrical field across the liquid-gas interface, the prompt
scintillation photons can be detected along with the
electroluminescence in the gas, produced by the ionization electrons
drifted to and extracted from the liquid surface. This technique
provides excellent vertex reconstruction, enabling a powerful target
fiducialization and a good discrimination between the nuclear recoil
signals and the electron recoil background~\cite{Aprile:2009dv}.

At present, there is a tight ongoing race between a few xenon
experiments. LUX~\cite{Akerib:2012ys}, a 250-kg xenon experiment,
located in the Sanford Underground Research Facility
(SURF)~\cite{Mei:2005gm}, USA, started taking physics data in 2013,
and recently concluded in May 2016. By combining the 95 live days of
data taken in 2013 and another 332 live days of data taken from 2014
to 2016, the collaboration reported a minimum of $1.1\times10^{-46}$
cm$^2$ upper limit at a WIMP mass of 50 GeV/c$^2$ on the WIMP-nucleon
cross-section~\cite{Akerib:2016vxi}. This is the strongest reported
limit to date. The best published WIMP search limit is set by the
half-ton scale PandaX-II~\cite{Tan:2016diz} experiment located in the
JinPing underground Laboratory (CJPL)~\cite{Kang:2010zza} in
China. This is also the largest running dual-phase xenon detector in
the world. The results were obtained with an exposure of
$3.3\times10^4$ kg-day, with an unprecedented background level of
$2\times10^{-3}$ events kg$^{-1}$d$^{-1}$keV$^{-1}$ within the
electron equivalent energy region between 1.3 and 8.7 keV.  The lowest
limit for the cross-section was set at $2.5\times10^{-46}$ cm$^2$ at a
WIMP mass of 40 GeV/c$^2$ (ref.~\cite{Tan:2016zwf}). PandaX-II will
continue data taking until 2018. The XENON100 experiment located in
the Gran Sasso National Laboratory (LNGS) in Italy recently reported
their final WIMP results using a total of 477 live days of data with a
limit of $1.1\times10^{-45}$ cm$^2$ at 50 GeV/c$^2$
(ref.~\cite{Aprile:2016swn}).  For these detectors at the
hundreds-kilogram level, the background due to uniformly distributed
sources such as $^{85}$Kr, $^{222}$Rn and $^{220}$Rn already becomes
prominent, and the effective suppression of such a background for the
next generation of experiments is under intensive research and
development. Looking into the near future, the
XENON1T~\cite{Aprile:2015uzo} experiment, the successor of XENON100,
is commissioning its dual-phase detector with a 2-ton liquid-xenon
target. The minimum projected sensitivity on the WIMP-nucleon
cross-section can reach $2.0\times10^{-47}$ cm$^2$ at 50 GeV/c$^2$
with an exposure of 2.2 ton-year~\cite{Aprile:2015uzo}. The upgrade
experiment XENONnT, with a 6.5-ton liquid-xenon target, is in
planning. The successor of LUX, LZ~\cite{Akerib:2015cja}, will contain
a 7-ton liquid-xenon target, and is expected to start operation in
2020 to achieve a sensitivity of $3\times10^{-48}$ cm$^2$ at 40
GeV/c$^2$ with 1,000 live days of data. In China, a future 4-ton scale
experiment, PandaX-4T, and its follow-up PandaX-30T are planned in the
second phase of CJPL. Another future large direct detection project,
DARWIN, is aiming at a target mass up to
40-ton~\cite{Aalbers:2016jon}.

The XMASS~\cite{Minamino:2010zz} experiment, in the Kamiokamine in
Japan, uses an alternative single-phase liquid-xenon detector to search
for WIMPs by detecting only the scintillation
photons~\cite{Mei:2005gm}. The detector has an approximately spherical
shape with a nearly 4$\pi$ photodetector coverage. One major challenge
for such a detector is to correctly reconstruct event vertices to have
an unambiguous fiducial volume selection. After discovering a
significant background from the photodetectors, which could `leak'
into the central region due to mis-reconstruction, the photodetector
array was replaced and the experiment resumed data taking in
2014~\cite{Abe:2016wqw}. The next stage for XMASS is XMASS 1.5
(ref.~\cite{Ichimura:2015xqs}), a detector with 5 tons of xenon in the
full volume. The projected sensitivity will reach 10$^{-46}$ cm$^{2}$
for 100GeV/c$^2$ WIMPs.

Argon is another type of noble element widely used in direct detection
experiments with a primary focus on the high-mass WIMPs. In comparison
to xenon, due to its low cost, even a few hundred-ton future detector
is foreseeable. The DarkSide-50 (ref.~\cite{Wright:2012raa})
experiment located in LNGS is running a dual-phase argon detector. The
first physical result of DarkSide-50 with atmospheric argon was
reported in 2015~\cite{Agnes:2014bvk}. It demonstrated excellent
pulseshape discrimination of the background electron recoil
events. For detectors using natural argon, $^{39}$Ar is an irreducible
long-lived cosmogenic background. To tame this background, DarkSide
collaboration pioneered the underground argon extraction and showed
that the $^{39}$Ar levels can be reduced by a factor of
$1.4\times10^3$ (ref.~\cite{Agnes:2015ftt}). At the next stage,
DarkSide-20k (ref.~\cite{Davini:2016vpd}), a detector with 20-ton of
argon, is expected to reach a sensitivity of $9\times10^{-48}$ cm$^2$
at 1 TeV/c$^2$ with a 100 ton-year exposure.  DEAP-3600
(ref.~\cite{Amaudruz:2014nsa}) at SNOLAB~\cite{Mei:2005gm} in Sudbury,
Canada is another argon-based experiment with a 1-ton target.  It is
operating in the single-liquid phase by detecting scintillation
photons. Pulseshape discrimination will be used to identify the
electron recoil background, which from a smaller prototype
detector~\cite{Amaudruz:2016qqa} has been proven to have less than
$2.7\times10^{-8}$ background contamination. Presently, DEAP-3600 is
being commissioned~\cite{Fatemighomi:2016ree}. Like DarkSide,
DEAP-3600 aims to eventually use underground argon extraction to
suppress $^{39}$Ar. The projected sensitivity on the spin-independent
WIMP-nucleon scattering cross-section can reach $10^{-46}$ cm$^2$ for
WIMP masses of 100 GeV/c$^2$ with a 3 ton-year
exposure~\cite{Amaudruz:2014nsa}.

An important frontier is the search for lighter WIMPs in the range of
a few GeV/c$^2$ which would produce a lower recoil energy. Aside from
the standard background issues common to all direct searches, a key
issue in such experiments is the operation of the detectors at a very
low threshold, hundreds of eV or lower. The
SuperCDMS~\cite{Agnese:2014aze} experiment located in the
Soudan~\cite{Mei:2005gm} mine in Minnesota, USA uses cryogenic
semiconductor detectors, armed with the so-called interleaved
Z-sensitive ionization phonon (iZIP) technique to detect both the
phonon and ionization signals at a low temperature of $\sim$40-50
mK. The technology provides a factor of greater than $10^6$ rejection of
the electron recoil background relative to the nuclear recoil
background. The collaboration published results in 2014 with 15 iZIP
modules, each with a mass of 0.6 kg (ref.~\cite{Agnese:2014aze}). To
reach a lower threshold, one of the iZIP detectors was operated with a
relatively high bias voltage (HV) to convert the ionization signal
into phonons, which reached an electron recoil threshold of 56
eV. With a 70 kg-day exposure, the experiment set the leading limits
published on low-mass WIMPs between 1.6 and 5.5 GeV/c$^2$
(ref.~\cite{Agnese:2015nto}). The CDEX~\cite{Kang:2013sjq} experiment
located in CJPL used point-contact germanium detectors operating at
liquid nitrogen temperature. These detectors have also the advantage
of a low threshold and a good rejection power to surface background,
and hence are suitable for low-mass WIMP searches. The first stage
CDEX-I experiment was completed with a 915-g detector at 475 eV
threshold with final exposure of 335.6
kg-days~\cite{Zhao:2016dak}. With no sign of unusual events, the
CDEX-I results challenged the WIMP interpretation of the event excess
from the CoGENT experiment which pioneered the
technology~\cite{Aalseth:2010vx}. An improved 1-kg CDEX detector with
an even lower threshold is currently taking data.

The DAMIC~\cite{Barreto:2011zu} experiment is another direct dark
matter search at SNOLAB. It uses high-resistivity charge-coupled
detectors (silicon) to record both the amplitude and position of the
ionization signal created by the nuclear recoil. The experiment is
particularly sensitive to WIMPs with low mass in the range 1$\sim$20
GeV/c$^2$. Results from the 0.6 kg-day exposure data taken in
2016~\cite{Aguilar-Arevalo:2016ndq} provided an upper limit for the
elastic spin-independent WIMP-nucleon cross-section below 10$^{-39}$
cm$^2$ for WIMPs with masses larger than 3 GeV/c$^2$.

Another low-mass WIMP search experiment is
CRESST~\cite{Bravin:1999fc}, located in LNGS. CRESST uses CaWO$_4$
crystal modules operated at an extremely low temperature (10 mK) to
detect both thermal excitations and scintillations resulting from the
nuclear recoils. The CRESST-II experiment, which concluded in 2015, is
currently leading the exclusion limit below a mass of 1.5 GeV/c$^2$,
and has extended below 1 GeV/c$^2$ for the first
time~\cite{Angloher:2015ewa}.

Most of the projects discussed above have planned upgrades to improve
their sensitivities,. At the next stage SuperCDMS will be moved to
SNOLAB, and it is projected to start operation in 2020. With an
exposure of 100 kg-year of Ge and 14.4 kg-year of Si with mixed iZIP
and HV mode, the SuperCDMS experiment is anticipated to approach the
ultimate neutrino background for WIMP masses in the range between 0.5
and 7 GeV/c$^2$ (ref.~\cite{Agnese:2016cpb}). Also at SNOLAB, the
upgraded DAMIC100 experiment with 100 g of target mass is upcoming. In
China, the CDEX-10 with a total target mass of 10 kg is being prepared
with an improved low-energy threshold. At LNGS, the newly designed
CRESST-III experiment started commissioning in June 2016, and a low
threshold below 100 eV is expected~\cite{Strauss:2016sxp}. In fact,
such lower threshold experiments are opening up new opportunities to
detect sub-GeV WIMPs via the very low energy electron recoil signals
in addition to the nuclear recoil signals. Besides existing
technologies, new experimental concepts are being discussed and
developed~\cite{Alexander:2016aln}.

\begin{figure}[htbp]
  \centering
  \includegraphics[width=0.95\linewidth]{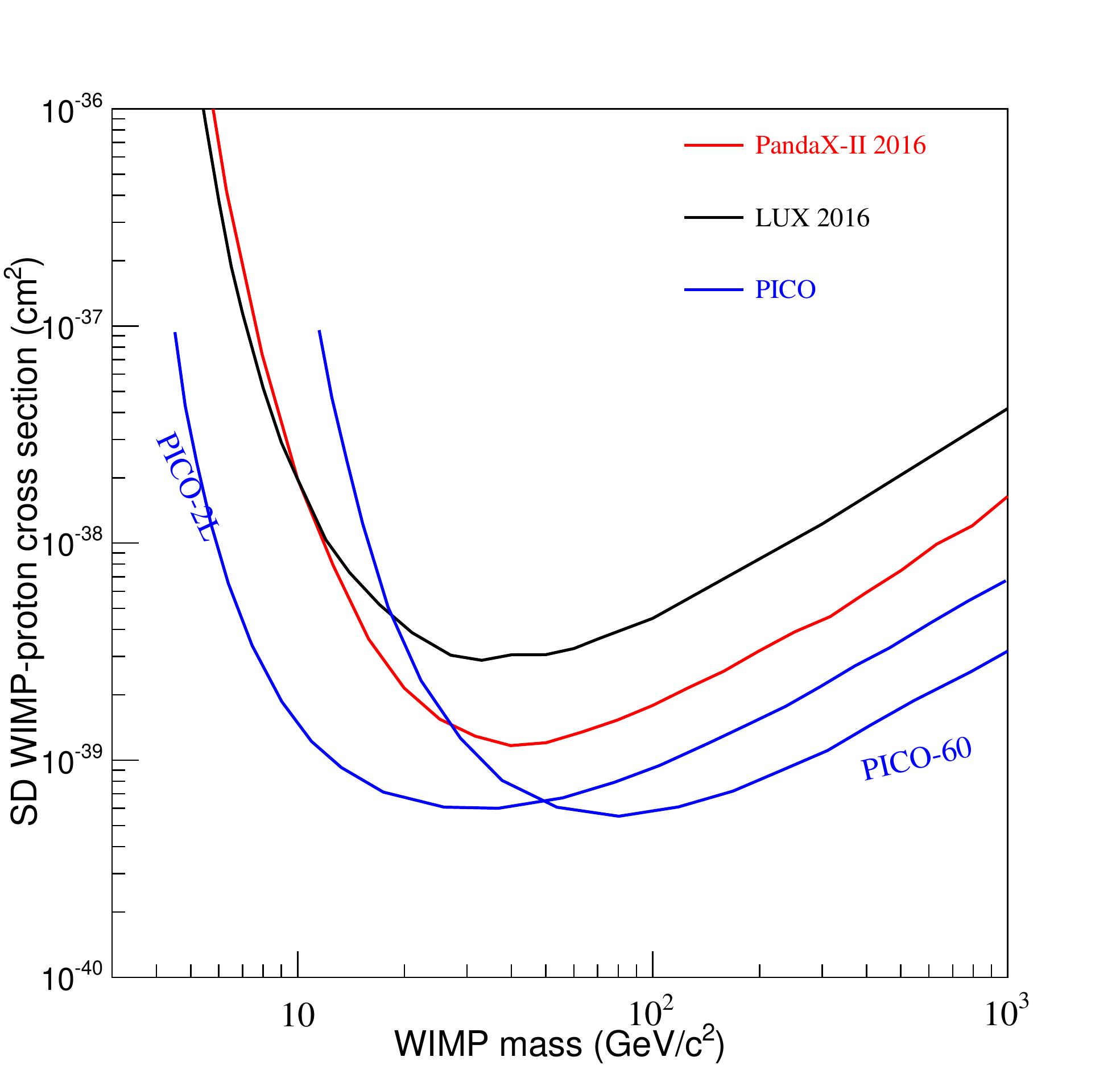}
  \caption{Upper limits on the spin-dependent (SD) WIMP-proton
    scattering cross-section set by diferent experiments. The limit
    curves are from LUX~\cite{Akerib:2016lao},
    PandaX-II~\cite{Fu:2016ega}, and PICO~\cite{Amole:2015pla,
      Amole:2016pye}.}
  \label{fig:sd_proton}
\end{figure}

\begin{figure}[htbp]
  \centering
  \includegraphics[width=0.95\linewidth]{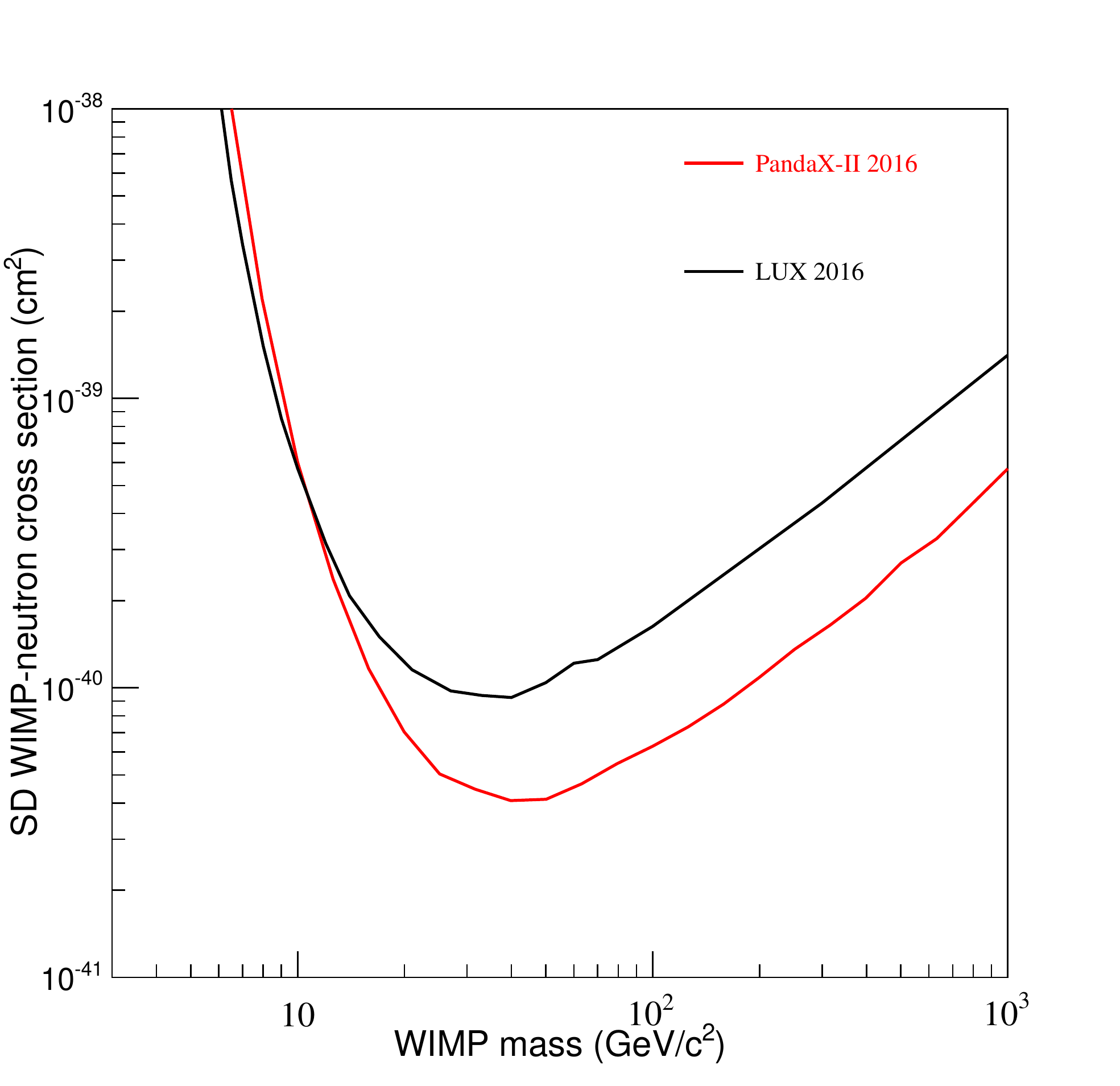}
  \caption{Upper limits on the spin-dependent WIMP-neutron scattering
    cross-section set by diferent xenon-based experiments. Limit
    curves from LUX~\cite{Akerib:2016lao} and
    PandaX-II~\cite{Fu:2016ega}.}
  \label{fig:sd_neutron}
\end{figure}

Despite the heated competition in searching for signs of
spin-independent (scalar) WIMP-nucleon scattering, WIMPs could well
carry spin and interact with nucleons through spin-dependent
interactions. At SNOLAB, the PICO~\cite{Amole:2015lsj} experiment is
looking for spin-dependent WIMPs signals with the bubble chamber
technique. In such superheated liquid detectors, only nuclear recoil
events with large enough stopping power can produce nucleation to
critically sized bubbles, which can then be photographed. Currently,
PICO is operating two chambers. One is called PICO-2L with 2.9 kg of
C$_3$F$_8$. Another one, PICO-60 (ref.~\cite{Amole:2015pla}), is the
largest bubble chamber used in dark matter search to date, and was
filled with 36.8 kg of CF$_3$I in run 1 and C$_3$F$_8$ in run 2. Due
to the odd number of protons in $^19$F and the fact that the last
unpaired proton dominates the overall spin of $^{19}$F, PICO has
excellent sensitivity to spin-dependent WIMP-proton scattering. In
fact, the most recent result of PICO-2L provided the most stringent
direct detection constraints on such cross-sections for WIMP masses of
less than 50 GeV/c$^2$(ref.~\cite{Amole:2016pye}). The leading
experiments searching for the WIMP-neutron interaction are again using
liquid xenon, relying on the fact that two of its natural isotopes,
$^{129}$Xe and $^{131}$Xe, carry odd numbers of neutrons. The lowest
published cross-section limit is $9.4\times10^{-41}$ cm$^2$ at 33
GeV/c$^2$ from the LUX collaboration with a total of $1.14\times10^4$
kg-day exposure~\cite{Akerib:2016lao}.  More recently, the PandaX-II
experiment reported a record limit of $4.1\times10^{-41}$ cm$^2$ at 40
GeV/c$^2$ in a preprint~\cite{Fu:2016ega}. The current limits set on
spin-dependent WIMP-nucleon scattering cross-sections by different
experiments are shown in Fig.~\ref{fig:sd_proton} (proton) and
Fig.~\ref{fig:sd_neutron} (neutron).

Finally, to prove the astrophysical nature of the WIMPs requires
measurements of the angular correlation between the recoil signals and
the galactic rotation. Relevant experimental techniques can also be
applied to further reject both electron and nuclear recoil
backgrounds, or even break the ultimate neutrino background
limit. Active research and development is being pursued globally in
the direction of solid and gaseous detectors, to demonstrate their
ability to identify recoil tracks and the scalability to a significant
target mass~\cite{Mayet:2016zxu}.

\begin{figure}[htbp]
  \centering
  \includegraphics[width=0.95\linewidth]{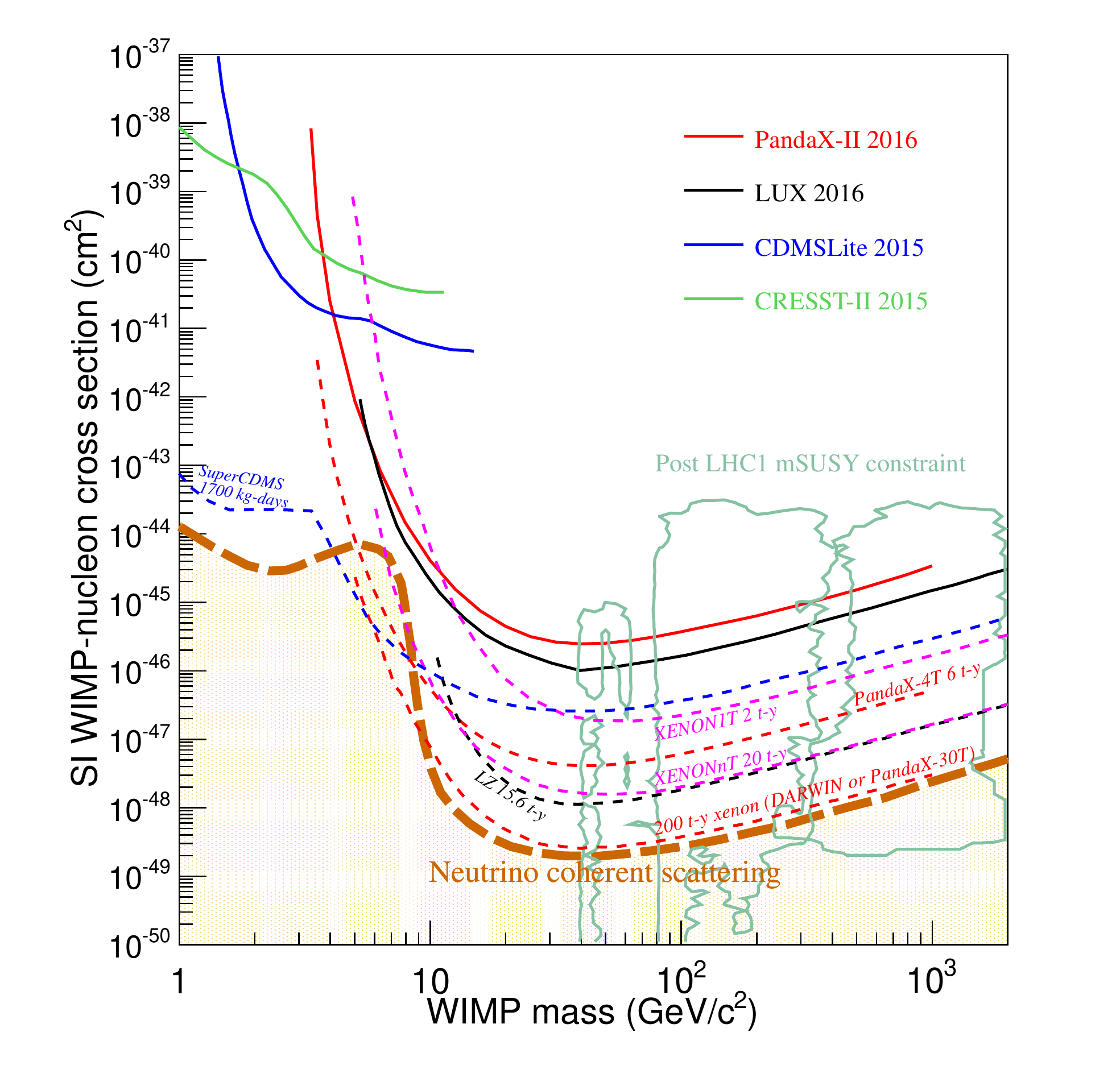}
  \caption{The projected sensitivity (dashed curves) on the
    spin-independent WIMP-nucleon cross-sections of a selected number
    of upcoming and planned direct detection experiments, including
    XENON1T~\cite{Aprile:2015uzo}, PandaX-4T,
    XENONnT~\cite{Aprile:2015uzo}, LZ~\cite{Akerib:2015cja},
    DARWIN~\cite{Aalbers:2016jon} or PandaX-30T, and
    SuperCDMS~\cite{Agnese:2016cpb}. Currently leading limits in
    Fig.~\ref{fig:limit_si_current} (see legend), the neutrino
    `floor'~\cite{Billard:2013qya}, and the post-LHC-Run1 minimal-SUSY
    allowed contours~\cite{Bagnaschi:2015eha} are overlaid in solid curves for
    comparison. The different crossings of the experimental
    sensitivities and the neutrino floor at around a few GeV/c$^2$ are
    primarily due to different threshold assumptions.  }
  \label{fig:si_projected}
\end{figure}

To summarize, we presented a non-exhaustive survey of the recent
progress in direct detection WIMP experiments. Although there has been
no solid evidence of a real event yet, the currently operating
experiments are expected to enhance their search sensitivity, and may
have the opportunity to first detect a real WIMP signal. In the next
decade or so, future experiments are planning to push the search
sensitivity in spin-independent WIMP-nucleon interaction to the
irreducible neutrino background in almost the entire WIMP range. The
present status and future reach of this very competitive field is
illustrated in Fig.~\ref{fig:si_projected}. Should a future
observation be made in one experiment, a robust discovery would
require confirmation from other experiments, preferably with different
experimental techniques and different target materials, as well as
cross checks from indirect and collider searches (for example, see
SUSY contours from Figs~\ref{fig:limit_si_current} and
~\ref{fig:si_projected}). This calls strongly for a worldwide
multi-faceted programme for dark matter detection. Finally, one cannot
ignore the importance of those null searches which have been setting
tighter constraints to many theoretical models and which may
eventually direct us on a completely different path towards
understanding this mysterious component of our Universe.

\begin{acknowledgments}
  This work is supported by grants from the National Science
  Foundation of China (Nos. 11435008, 11455001, 11505112 and
  11525522), a grant from the Ministry of Science and Technology of
  China (Grant No. 2016YFA0400301), and in part by the Chinese Academy
  of Sciences Center for Excellence in Particle Physics (CCEPP), the
  Key Laboratory for Particle Physics, Astrophysics and Cosmology,
  Ministry of Education, and Shanghai Key Laboratory for Particle
  Physics and Cosmology (SKLPPC). Finally, we thank the Hong Kong
  Hongwen Foundation for financial support.
\end{acknowledgments}

\bibliographystyle{apsrev4-1}
\bibliography{refs.bib}

\begin{thebibliography}{64}%
\makeatletter
\providecommand \@ifxundefined [1]{%
 \@ifx{#1\undefined}
}%
\providecommand \@ifnum [1]{%
 \ifnum #1\expandafter \@firstoftwo
 \else \expandafter \@secondoftwo
 \fi
}%
\providecommand \@ifx [1]{%
 \ifx #1\expandafter \@firstoftwo
 \else \expandafter \@secondoftwo
 \fi
}%
\providecommand \natexlab [1]{#1}%
\providecommand \enquote  [1]{``#1''}%
\providecommand \bibnamefont  [1]{#1}%
\providecommand \bibfnamefont [1]{#1}%
\providecommand \citenamefont [1]{#1}%
\providecommand \href@noop [0]{\@secondoftwo}%
\providecommand \href [0]{\begingroup \@sanitize@url \@href}%
\providecommand \@href[1]{\@@startlink{#1}\@@href}%
\providecommand \@@href[1]{\endgroup#1\@@endlink}%
\providecommand \@sanitize@url [0]{\catcode `\\12\catcode `\$12\catcode
  `\&12\catcode `\#12\catcode `\^12\catcode `\_12\catcode `\%12\relax}%
\providecommand \@@startlink[1]{}%
\providecommand \@@endlink[0]{}%
\providecommand \url  [0]{\begingroup\@sanitize@url \@url }%
\providecommand \@url [1]{\endgroup\@href {#1}{\urlprefix }}%
\providecommand \urlprefix  [0]{URL }%
\providecommand \Eprint [0]{\href }%
\providecommand \doibase [0]{http://dx.doi.org/}%
\providecommand \selectlanguage [0]{\@gobble}%
\providecommand \bibinfo  [0]{\@secondoftwo}%
\providecommand \bibfield  [0]{\@secondoftwo}%
\providecommand \translation [1]{[#1]}%
\providecommand \BibitemOpen [0]{}%
\providecommand \bibitemStop [0]{}%
\providecommand \bibitemNoStop [0]{.\EOS\space}%
\providecommand \EOS [0]{\spacefactor3000\relax}%
\providecommand \BibitemShut  [1]{\csname bibitem#1\endcsname}%
\let\auto@bib@innerbib\@empty
\bibitem [{\citenamefont {Bertone}\ \emph {et~al.}(2005)\citenamefont
  {Bertone}, \citenamefont {Hooper},\ and\ \citenamefont
  {Silk}}]{Bertone:2004pz}%
  \BibitemOpen
  \bibfield  {author} {\bibinfo {author} {\bibfnamefont {G.}~\bibnamefont
  {Bertone}}, \bibinfo {author} {\bibfnamefont {D.}~\bibnamefont {Hooper}}, \
  and\ \bibinfo {author} {\bibfnamefont {J.}~\bibnamefont {Silk}},\ }\href
  {\doibase 10.1016/j.physrep.2004.08.031} {\bibfield  {journal} {\bibinfo
  {journal} {Phys. Rept.}\ }\textbf {\bibinfo {volume} {405}},\ \bibinfo
  {pages} {279} (\bibinfo {year} {2005})},\ \Eprint
  {http://arxiv.org/abs/hep-ph/0404175} {arXiv:hep-ph/0404175 [hep-ph]}
  \BibitemShut {NoStop}%
\bibitem [{\citenamefont {Savage}\ \emph {et~al.}(2006)\citenamefont {Savage},
  \citenamefont {Freese},\ and\ \citenamefont {Gondolo}}]{Savage:2006qr}%
  \BibitemOpen
  \bibfield  {author} {\bibinfo {author} {\bibfnamefont {C.}~\bibnamefont
  {Savage}}, \bibinfo {author} {\bibfnamefont {K.}~\bibnamefont {Freese}}, \
  and\ \bibinfo {author} {\bibfnamefont {P.}~\bibnamefont {Gondolo}},\ }\href
  {\doibase 10.1103/PhysRevD.74.043531} {\bibfield  {journal} {\bibinfo
  {journal} {Phys. Rev.}\ }\textbf {\bibinfo {volume} {D74}},\ \bibinfo {pages}
  {043531} (\bibinfo {year} {2006})},\ \Eprint
  {http://arxiv.org/abs/astro-ph/0607121} {arXiv:astro-ph/0607121 [astro-ph]}
  \BibitemShut {NoStop}%
\bibitem [{\citenamefont {Jungman}\ \emph {et~al.}(1996)\citenamefont
  {Jungman}, \citenamefont {Kamionkowski},\ and\ \citenamefont
  {Griest}}]{Jungman:1995df}%
  \BibitemOpen
  \bibfield  {author} {\bibinfo {author} {\bibfnamefont {G.}~\bibnamefont
  {Jungman}}, \bibinfo {author} {\bibfnamefont {M.}~\bibnamefont
  {Kamionkowski}}, \ and\ \bibinfo {author} {\bibfnamefont {K.}~\bibnamefont
  {Griest}},\ }\href {\doibase 10.1016/0370-1573(95)00058-5} {\bibfield
  {journal} {\bibinfo  {journal} {Phys. Rept.}\ }\textbf {\bibinfo {volume}
  {267}},\ \bibinfo {pages} {195} (\bibinfo {year} {1996})},\ \Eprint
  {http://arxiv.org/abs/hep-ph/9506380} {arXiv:hep-ph/9506380 [hep-ph]}
  \BibitemShut {NoStop}%
\bibitem [{\citenamefont {Smith}\ \emph {et~al.}(2007)\citenamefont {Smith}
  \emph {et~al.}}]{Smith:2006ym}%
  \BibitemOpen
  \bibfield  {author} {\bibinfo {author} {\bibfnamefont {M.~C.}\ \bibnamefont
  {Smith}} \emph {et~al.},\ }\href {\doibase 10.1111/j.1365-2966.2007.11964.x}
  {\bibfield  {journal} {\bibinfo  {journal} {Mon. Not. Roy. Astron. Soc.}\
  }\textbf {\bibinfo {volume} {379}},\ \bibinfo {pages} {755} (\bibinfo {year}
  {2007})},\ \Eprint {http://arxiv.org/abs/astro-ph/0611671}
  {arXiv:astro-ph/0611671 [astro-ph]} \BibitemShut {NoStop}%
\bibitem [{\citenamefont {Peccei}\ and\ \citenamefont
  {Quinn}(1977)}]{Peccei:1977hh}%
  \BibitemOpen
  \bibfield  {author} {\bibinfo {author} {\bibfnamefont {R.~D.}\ \bibnamefont
  {Peccei}}\ and\ \bibinfo {author} {\bibfnamefont {H.~R.}\ \bibnamefont
  {Quinn}},\ }\href {\doibase 10.1103/PhysRevLett.38.1440} {\bibfield
  {journal} {\bibinfo  {journal} {Phys. Rev. Lett.}\ }\textbf {\bibinfo
  {volume} {38}},\ \bibinfo {pages} {1440} (\bibinfo {year}
  {1977})}\BibitemShut {NoStop}%
\bibitem [{\citenamefont {Wilczek}(1978)}]{Wilczek:1977pj}%
  \BibitemOpen
  \bibfield  {author} {\bibinfo {author} {\bibfnamefont {F.}~\bibnamefont
  {Wilczek}},\ }\href {\doibase 10.1103/PhysRevLett.40.279} {\bibfield
  {journal} {\bibinfo  {journal} {Phys. Rev. Lett.}\ }\textbf {\bibinfo
  {volume} {40}},\ \bibinfo {pages} {279} (\bibinfo {year} {1978})}\BibitemShut
  {NoStop}%
\bibitem [{\citenamefont {Kim}(1987)}]{Kim:1986ax}%
  \BibitemOpen
  \bibfield  {author} {\bibinfo {author} {\bibfnamefont {J.~E.}\ \bibnamefont
  {Kim}},\ }\href {\doibase 10.1016/0370-1573(87)90017-2} {\bibfield  {journal}
  {\bibinfo  {journal} {Phys. Rept.}\ }\textbf {\bibinfo {volume} {150}},\
  \bibinfo {pages} {1} (\bibinfo {year} {1987})}\BibitemShut {NoStop}%
\bibitem [{\citenamefont {Marsh}(2016)}]{Marsh:2015xka}%
  \BibitemOpen
  \bibfield  {author} {\bibinfo {author} {\bibfnamefont {D.~J.~E.}\
  \bibnamefont {Marsh}},\ }\href {\doibase 10.1016/j.physrep.2016.06.005}
  {\bibfield  {journal} {\bibinfo  {journal} {Phys. Rept.}\ }\textbf {\bibinfo
  {volume} {643}},\ \bibinfo {pages} {1} (\bibinfo {year} {2016})},\ \Eprint
  {http://arxiv.org/abs/1510.07633} {arXiv:1510.07633 [astro-ph.CO]}
  \BibitemShut {NoStop}%
\bibitem [{\citenamefont {Gaskins}(2016)}]{Gaskins:2016cha}%
  \BibitemOpen
  \bibfield  {author} {\bibinfo {author} {\bibfnamefont {J.~M.}\ \bibnamefont
  {Gaskins}},\ }\href {\doibase 10.1080/00107514.2016.1175160} {\bibfield
  {journal} {\bibinfo  {journal} {Contemp. Phys.}\ }\textbf {\bibinfo {volume}
  {57}},\ \bibinfo {pages} {496} (\bibinfo {year} {2016})},\ \Eprint
  {http://arxiv.org/abs/1604.00014} {arXiv:1604.00014 [astro-ph.HE]}
  \BibitemShut {NoStop}%
\bibitem [{\citenamefont {Lewin}\ and\ \citenamefont
  {Smith}(1996)}]{Lewin:1995rx}%
  \BibitemOpen
  \bibfield  {author} {\bibinfo {author} {\bibfnamefont {J.~D.}\ \bibnamefont
  {Lewin}}\ and\ \bibinfo {author} {\bibfnamefont {P.~F.}\ \bibnamefont
  {Smith}},\ }\href {\doibase 10.1016/S0927-6505(96)00047-3} {\bibfield
  {journal} {\bibinfo  {journal} {Astropart. Phys.}\ }\textbf {\bibinfo
  {volume} {6}},\ \bibinfo {pages} {87} (\bibinfo {year} {1996})}\BibitemShut
  {NoStop}%
\bibitem [{\citenamefont {Strigari}(2009)}]{Strigari:2009bq}%
  \BibitemOpen
  \bibfield  {author} {\bibinfo {author} {\bibfnamefont {L.~E.}\ \bibnamefont
  {Strigari}},\ }\href {\doibase 10.1088/1367-2630/11/10/105011} {\bibfield
  {journal} {\bibinfo  {journal} {New J. Phys.}\ }\textbf {\bibinfo {volume}
  {11}},\ \bibinfo {pages} {105011} (\bibinfo {year} {2009})},\ \Eprint
  {http://arxiv.org/abs/0903.3630} {arXiv:0903.3630 [astro-ph.CO]} \BibitemShut
  {NoStop}%
\bibitem [{\citenamefont {Gutlein}\ \emph {et~al.}(2010)\citenamefont {Gutlein}
  \emph {et~al.}}]{Gutlein:2010tq}%
  \BibitemOpen
  \bibfield  {author} {\bibinfo {author} {\bibfnamefont {A.}~\bibnamefont
  {Gutlein}} \emph {et~al.},\ }\href {\doibase
  10.1016/j.astropartphys.2010.06.002} {\bibfield  {journal} {\bibinfo
  {journal} {Astropart. Phys.}\ }\textbf {\bibinfo {volume} {34}},\ \bibinfo
  {pages} {90} (\bibinfo {year} {2010})},\ \Eprint
  {http://arxiv.org/abs/1003.5530} {arXiv:1003.5530 [hep-ph]} \BibitemShut
  {NoStop}%
\bibitem [{\citenamefont {Ruppin}\ \emph {et~al.}(2014)\citenamefont {Ruppin},
  \citenamefont {Billard}, \citenamefont {Figueroa-Feliciano},\ and\
  \citenamefont {Strigari}}]{Ruppin:2014bra}%
  \BibitemOpen
  \bibfield  {author} {\bibinfo {author} {\bibfnamefont {F.}~\bibnamefont
  {Ruppin}}, \bibinfo {author} {\bibfnamefont {J.}~\bibnamefont {Billard}},
  \bibinfo {author} {\bibfnamefont {E.}~\bibnamefont {Figueroa-Feliciano}}, \
  and\ \bibinfo {author} {\bibfnamefont {L.}~\bibnamefont {Strigari}},\ }\href
  {\doibase 10.1103/PhysRevD.90.083510} {\bibfield  {journal} {\bibinfo
  {journal} {Phys. Rev.}\ }\textbf {\bibinfo {volume} {D90}},\ \bibinfo {pages}
  {083510} (\bibinfo {year} {2014})},\ \Eprint {http://arxiv.org/abs/1408.3581}
  {arXiv:1408.3581 [hep-ph]} \BibitemShut {NoStop}%
\bibitem [{\citenamefont {Dedes}\ \emph {et~al.}(2010)\citenamefont {Dedes},
  \citenamefont {Giomataris}, \citenamefont {Suxho},\ and\ \citenamefont
  {Vergados}}]{Dedes:2009bk}%
  \BibitemOpen
  \bibfield  {author} {\bibinfo {author} {\bibfnamefont {A.}~\bibnamefont
  {Dedes}}, \bibinfo {author} {\bibfnamefont {I.}~\bibnamefont {Giomataris}},
  \bibinfo {author} {\bibfnamefont {K.}~\bibnamefont {Suxho}}, \ and\ \bibinfo
  {author} {\bibfnamefont {J.~D.}\ \bibnamefont {Vergados}},\ }\href {\doibase
  10.1016/j.nuclphysb.2009.09.032} {\bibfield  {journal} {\bibinfo  {journal}
  {Nucl. Phys.}\ }\textbf {\bibinfo {volume} {B826}},\ \bibinfo {pages} {148}
  (\bibinfo {year} {2010})},\ \Eprint {http://arxiv.org/abs/0907.0758}
  {arXiv:0907.0758 [hep-ph]} \BibitemShut {NoStop}%
\bibitem [{\citenamefont {Gaitskell}(2004)}]{Gaitskell:2004gd}%
  \BibitemOpen
  \bibfield  {author} {\bibinfo {author} {\bibfnamefont {R.~J.}\ \bibnamefont
  {Gaitskell}},\ }\href {\doibase 10.1146/annurev.nucl.54.070103.181244}
  {\bibfield  {journal} {\bibinfo  {journal} {Ann. Rev. Nucl. Part. Sci.}\
  }\textbf {\bibinfo {volume} {54}},\ \bibinfo {pages} {315} (\bibinfo {year}
  {2004})}\BibitemShut {NoStop}%
\bibitem [{\citenamefont {Bernabei}\ \emph {et~al.}(2013)\citenamefont
  {Bernabei} \emph {et~al.}}]{Bernabei:2013xsa}%
  \BibitemOpen
  \bibfield  {author} {\bibinfo {author} {\bibfnamefont {R.}~\bibnamefont
  {Bernabei}} \emph {et~al.},\ }\href {\doibase 10.1140/epjc/s10052-013-2648-7}
  {\bibfield  {journal} {\bibinfo  {journal} {Eur. Phys. J.}\ }\textbf
  {\bibinfo {volume} {C73}},\ \bibinfo {pages} {2648} (\bibinfo {year}
  {2013})},\ \Eprint {http://arxiv.org/abs/1308.5109} {arXiv:1308.5109
  [astro-ph.GA]} \BibitemShut {NoStop}%
\bibitem [{\citenamefont {Aalseth}\ \emph {et~al.}(2013)\citenamefont {Aalseth}
  \emph {et~al.}}]{Aalseth:2012if}%
  \BibitemOpen
  \bibfield  {author} {\bibinfo {author} {\bibfnamefont {C.~E.}\ \bibnamefont
  {Aalseth}} \emph {et~al.} (\bibinfo {collaboration} {CoGeNT}),\ }\href
  {\doibase 10.1103/PhysRevD.88.012002} {\bibfield  {journal} {\bibinfo
  {journal} {Phys. Rev.}\ }\textbf {\bibinfo {volume} {D88}},\ \bibinfo {pages}
  {012002} (\bibinfo {year} {2013})},\ \Eprint {http://arxiv.org/abs/1208.5737}
  {arXiv:1208.5737 [astro-ph.CO]} \BibitemShut {NoStop}%
\bibitem [{\citenamefont {Angloher}\ \emph {et~al.}(2012)\citenamefont
  {Angloher} \emph {et~al.}}]{Angloher:2011uu}%
  \BibitemOpen
  \bibfield  {author} {\bibinfo {author} {\bibfnamefont {G.}~\bibnamefont
  {Angloher}} \emph {et~al.},\ }\href {\doibase 10.1140/epjc/s10052-012-1971-8}
  {\bibfield  {journal} {\bibinfo  {journal} {Eur. Phys. J.}\ }\textbf
  {\bibinfo {volume} {C72}},\ \bibinfo {pages} {1971} (\bibinfo {year}
  {2012})},\ \Eprint {http://arxiv.org/abs/1109.0702} {arXiv:1109.0702
  [astro-ph.CO]} \BibitemShut {NoStop}%
\bibitem [{\citenamefont {Agnese}\ \emph {et~al.}(2013)\citenamefont {Agnese}
  \emph {et~al.}}]{Agnese:2013rvf}%
  \BibitemOpen
  \bibfield  {author} {\bibinfo {author} {\bibfnamefont {R.}~\bibnamefont
  {Agnese}} \emph {et~al.} (\bibinfo {collaboration} {CDMS}),\ }\href {\doibase
  10.1103/PhysRevLett.111.251301} {\bibfield  {journal} {\bibinfo  {journal}
  {Phys. Rev. Lett.}\ }\textbf {\bibinfo {volume} {111}},\ \bibinfo {pages}
  {251301} (\bibinfo {year} {2013})},\ \Eprint {http://arxiv.org/abs/1304.4279}
  {arXiv:1304.4279 [hep-ex]} \BibitemShut {NoStop}%
\bibitem [{\citenamefont {Billard}\ \emph {et~al.}(2014)\citenamefont
  {Billard}, \citenamefont {Strigari},\ and\ \citenamefont
  {Figueroa-Feliciano}}]{Billard:2013qya}%
  \BibitemOpen
  \bibfield  {author} {\bibinfo {author} {\bibfnamefont {J.}~\bibnamefont
  {Billard}}, \bibinfo {author} {\bibfnamefont {L.}~\bibnamefont {Strigari}}, \
  and\ \bibinfo {author} {\bibfnamefont {E.}~\bibnamefont
  {Figueroa-Feliciano}},\ }\href {\doibase 10.1103/PhysRevD.89.023524}
  {\bibfield  {journal} {\bibinfo  {journal} {Phys. Rev.}\ }\textbf {\bibinfo
  {volume} {D89}},\ \bibinfo {pages} {023524} (\bibinfo {year} {2014})},\
  \Eprint {http://arxiv.org/abs/1307.5458} {arXiv:1307.5458 [hep-ph]}
  \BibitemShut {NoStop}%
\bibitem [{\citenamefont {Bagnaschi}\ \emph {et~al.}(2015)\citenamefont
  {Bagnaschi} \emph {et~al.}}]{Bagnaschi:2015eha}%
  \BibitemOpen
  \bibfield  {author} {\bibinfo {author} {\bibfnamefont {E.~A.}\ \bibnamefont
  {Bagnaschi}} \emph {et~al.},\ }\href {\doibase
  10.1140/epjc/s10052-015-3718-9} {\bibfield  {journal} {\bibinfo  {journal}
  {Eur. Phys. J.}\ }\textbf {\bibinfo {volume} {C75}},\ \bibinfo {pages} {500}
  (\bibinfo {year} {2015})},\ \Eprint {http://arxiv.org/abs/1508.01173}
  {arXiv:1508.01173 [hep-ph]} \BibitemShut {NoStop}%
\bibitem [{\citenamefont {Angle}\ \emph {et~al.}(2008)\citenamefont {Angle}
  \emph {et~al.}}]{Angle:2007uj}%
  \BibitemOpen
  \bibfield  {author} {\bibinfo {author} {\bibfnamefont {J.}~\bibnamefont
  {Angle}} \emph {et~al.} (\bibinfo {collaboration} {XENON}),\ }\href {\doibase
  10.1103/PhysRevLett.100.021303} {\bibfield  {journal} {\bibinfo  {journal}
  {Phys. Rev. Lett.}\ }\textbf {\bibinfo {volume} {100}},\ \bibinfo {pages}
  {021303} (\bibinfo {year} {2008})},\ \Eprint {http://arxiv.org/abs/0706.0039}
  {arXiv:0706.0039 [astro-ph]} \BibitemShut {NoStop}%
\bibitem [{\citenamefont {Aprile}\ \emph {et~al.}(2013)\citenamefont {Aprile}
  \emph {et~al.}}]{Aprile:2013doa}%
  \BibitemOpen
  \bibfield  {author} {\bibinfo {author} {\bibfnamefont {E.}~\bibnamefont
  {Aprile}} \emph {et~al.} (\bibinfo {collaboration} {XENON100}),\ }\href
  {\doibase 10.1103/PhysRevLett.111.021301} {\bibfield  {journal} {\bibinfo
  {journal} {Phys. Rev. Lett.}\ }\textbf {\bibinfo {volume} {111}},\ \bibinfo
  {pages} {021301} (\bibinfo {year} {2013})},\ \Eprint
  {http://arxiv.org/abs/1301.6620} {arXiv:1301.6620 [astro-ph.CO]} \BibitemShut
  {NoStop}%
\bibitem [{\citenamefont {Akerib}\ \emph {et~al.}(2014)\citenamefont {Akerib}
  \emph {et~al.}}]{Akerib:2013tjd}%
  \BibitemOpen
  \bibfield  {author} {\bibinfo {author} {\bibfnamefont {D.~S.}\ \bibnamefont
  {Akerib}} \emph {et~al.} (\bibinfo {collaboration} {LUX}),\ }\href {\doibase
  10.1103/PhysRevLett.112.091303} {\bibfield  {journal} {\bibinfo  {journal}
  {Phys. Rev. Lett.}\ }\textbf {\bibinfo {volume} {112}},\ \bibinfo {pages}
  {091303} (\bibinfo {year} {2014})},\ \Eprint {http://arxiv.org/abs/1310.8214}
  {arXiv:1310.8214 [astro-ph.CO]} \BibitemShut {NoStop}%
\bibitem [{\citenamefont {Akerib}\ \emph
  {et~al.}(2015{\natexlab{a}})\citenamefont {Akerib} \emph
  {et~al.}}]{Akerib:2015rjg}%
  \BibitemOpen
  \bibfield  {author} {\bibinfo {author} {\bibfnamefont {D.~S.}\ \bibnamefont
  {Akerib}} \emph {et~al.} (\bibinfo {collaboration} {LUX}),\ }\href@noop {} {\
   (\bibinfo {year} {2015}{\natexlab{a}})},\ \Eprint
  {http://arxiv.org/abs/1512.03506} {arXiv:1512.03506 [astro-ph.CO]}
  \BibitemShut {NoStop}%
\bibitem [{\citenamefont {Tan}\ \emph {et~al.}(2016{\natexlab{a}})\citenamefont
  {Tan} \emph {et~al.}}]{Tan:2016zwf}%
  \BibitemOpen
  \bibfield  {author} {\bibinfo {author} {\bibfnamefont {A.}~\bibnamefont
  {Tan}} \emph {et~al.} (\bibinfo {collaboration} {PandaX-II}),\ }\href
  {\doibase 10.1103/PhysRevLett.117.121303} {\bibfield  {journal} {\bibinfo
  {journal} {Phys. Rev. Lett.}\ }\textbf {\bibinfo {volume} {117}},\ \bibinfo
  {pages} {121303} (\bibinfo {year} {2016}{\natexlab{a}})},\ \Eprint
  {http://arxiv.org/abs/1607.07400} {arXiv:1607.07400 [hep-ex]} \BibitemShut
  {NoStop}%
\bibitem [{\citenamefont {Aprile}\ and\ \citenamefont
  {Doke}(2010)}]{Aprile:2009dv}%
  \BibitemOpen
  \bibfield  {author} {\bibinfo {author} {\bibfnamefont {E.}~\bibnamefont
  {Aprile}}\ and\ \bibinfo {author} {\bibfnamefont {T.}~\bibnamefont {Doke}},\
  }\href {\doibase 10.1103/RevModPhys.82.2053} {\bibfield  {journal} {\bibinfo
  {journal} {Rev. Mod. Phys.}\ }\textbf {\bibinfo {volume} {82}},\ \bibinfo
  {pages} {2053} (\bibinfo {year} {2010})},\ \Eprint
  {http://arxiv.org/abs/0910.4956} {arXiv:0910.4956 [physics.ins-det]}
  \BibitemShut {NoStop}%
\bibitem [{\citenamefont {Akerib}\ \emph {et~al.}(2013)\citenamefont {Akerib}
  \emph {et~al.}}]{Akerib:2012ys}%
  \BibitemOpen
  \bibfield  {author} {\bibinfo {author} {\bibfnamefont {D.~S.}\ \bibnamefont
  {Akerib}} \emph {et~al.} (\bibinfo {collaboration} {LUX}),\ }\href {\doibase
  10.1016/j.nima.2012.11.135} {\bibfield  {journal} {\bibinfo  {journal} {Nucl.
  Instrum. Meth.}\ }\textbf {\bibinfo {volume} {A704}},\ \bibinfo {pages} {111}
  (\bibinfo {year} {2013})},\ \Eprint {http://arxiv.org/abs/1211.3788}
  {arXiv:1211.3788 [physics.ins-det]} \BibitemShut {NoStop}%
\bibitem [{\citenamefont {Mei}\ and\ \citenamefont {Hime}(2006)}]{Mei:2005gm}%
  \BibitemOpen
  \bibfield  {author} {\bibinfo {author} {\bibfnamefont {D.}~\bibnamefont
  {Mei}}\ and\ \bibinfo {author} {\bibfnamefont {A.}~\bibnamefont {Hime}},\
  }\href {\doibase 10.1103/PhysRevD.73.053004} {\bibfield  {journal} {\bibinfo
  {journal} {Phys. Rev.}\ }\textbf {\bibinfo {volume} {D73}},\ \bibinfo {pages}
  {053004} (\bibinfo {year} {2006})},\ \Eprint
  {http://arxiv.org/abs/astro-ph/0512125} {arXiv:astro-ph/0512125 [astro-ph]}
  \BibitemShut {NoStop}%
\bibitem [{\citenamefont {Akerib}\ \emph {et~al.}(2017)\citenamefont {Akerib}
  \emph {et~al.}}]{Akerib:2016vxi}%
  \BibitemOpen
  \bibfield  {author} {\bibinfo {author} {\bibfnamefont {D.~S.}\ \bibnamefont
  {Akerib}} \emph {et~al.} (\bibinfo {collaboration} {LUX}),\ }\href {\doibase
  10.1103/PhysRevLett.118.021303} {\bibfield  {journal} {\bibinfo  {journal}
  {Phys. Rev. Lett.}\ }\textbf {\bibinfo {volume} {118}},\ \bibinfo {pages}
  {021303} (\bibinfo {year} {2017})},\ \Eprint
  {http://arxiv.org/abs/1608.07648} {arXiv:1608.07648 [astro-ph.CO]}
  \BibitemShut {NoStop}%
\bibitem [{\citenamefont {Tan}\ \emph {et~al.}(2016{\natexlab{b}})\citenamefont
  {Tan} \emph {et~al.}}]{Tan:2016diz}%
  \BibitemOpen
  \bibfield  {author} {\bibinfo {author} {\bibfnamefont {A.}~\bibnamefont
  {Tan}} \emph {et~al.} (\bibinfo {collaboration} {PandaX}),\ }\href {\doibase
  10.1103/PhysRevD.93.122009} {\bibfield  {journal} {\bibinfo  {journal} {Phys.
  Rev.}\ }\textbf {\bibinfo {volume} {D93}},\ \bibinfo {pages} {122009}
  (\bibinfo {year} {2016}{\natexlab{b}})},\ \Eprint
  {http://arxiv.org/abs/1602.06563} {arXiv:1602.06563 [hep-ex]} \BibitemShut
  {NoStop}%
\bibitem [{\citenamefont {Kang}\ \emph {et~al.}(2010)\citenamefont {Kang},
  \citenamefont {Cheng}, \citenamefont {Chen}, \citenamefont {Li},
  \citenamefont {Shen}, \citenamefont {Wu},\ and\ \citenamefont
  {Yue}}]{Kang:2010zza}%
  \BibitemOpen
  \bibfield  {author} {\bibinfo {author} {\bibfnamefont {K.~J.}\ \bibnamefont
  {Kang}}, \bibinfo {author} {\bibfnamefont {J.~P.}\ \bibnamefont {Cheng}},
  \bibinfo {author} {\bibfnamefont {Y.~H.}\ \bibnamefont {Chen}}, \bibinfo
  {author} {\bibfnamefont {Y.~J.}\ \bibnamefont {Li}}, \bibinfo {author}
  {\bibfnamefont {M.~B.}\ \bibnamefont {Shen}}, \bibinfo {author}
  {\bibfnamefont {S.~Y.}\ \bibnamefont {Wu}}, \ and\ \bibinfo {author}
  {\bibfnamefont {Q.}~\bibnamefont {Yue}},\ }\bibfield  {booktitle} {\emph
  {\bibinfo {booktitle} {{Proceedings, 11th International Conference on Topics
  in astroparticle and underground physics in Memory of Julio Morales (TAUP
  2009): Rome, Italy, July 1-5, 2009}}},\ }\href {\doibase
  10.1088/1742-6596/203/1/012028} {\bibfield  {journal} {\bibinfo  {journal}
  {J. Phys. Conf. Ser.}\ }\textbf {\bibinfo {volume} {203}},\ \bibinfo {pages}
  {012028} (\bibinfo {year} {2010})}\BibitemShut {NoStop}%
\bibitem [{\citenamefont {Aprile}\ \emph
  {et~al.}(2016{\natexlab{a}})\citenamefont {Aprile} \emph
  {et~al.}}]{Aprile:2016swn}%
  \BibitemOpen
  \bibfield  {author} {\bibinfo {author} {\bibfnamefont {E.}~\bibnamefont
  {Aprile}} \emph {et~al.} (\bibinfo {collaboration} {XENON100}),\ }\href
  {\doibase 10.1103/PhysRevD.94.122001} {\bibfield  {journal} {\bibinfo
  {journal} {Phys. Rev.}\ }\textbf {\bibinfo {volume} {D94}},\ \bibinfo {pages}
  {122001} (\bibinfo {year} {2016}{\natexlab{a}})},\ \Eprint
  {http://arxiv.org/abs/1609.06154} {arXiv:1609.06154 [astro-ph.CO]}
  \BibitemShut {NoStop}%
\bibitem [{\citenamefont {Aprile}\ \emph
  {et~al.}(2016{\natexlab{b}})\citenamefont {Aprile} \emph
  {et~al.}}]{Aprile:2015uzo}%
  \BibitemOpen
  \bibfield  {author} {\bibinfo {author} {\bibfnamefont {E.}~\bibnamefont
  {Aprile}} \emph {et~al.} (\bibinfo {collaboration} {XENON}),\ }\href
  {\doibase 10.1088/1475-7516/2016/04/027} {\bibfield  {journal} {\bibinfo
  {journal} {JCAP}\ }\textbf {\bibinfo {volume} {1604}},\ \bibinfo {pages}
  {027} (\bibinfo {year} {2016}{\natexlab{b}})},\ \Eprint
  {http://arxiv.org/abs/1512.07501} {arXiv:1512.07501 [physics.ins-det]}
  \BibitemShut {NoStop}%
\bibitem [{\citenamefont {Akerib}\ \emph
  {et~al.}(2015{\natexlab{b}})\citenamefont {Akerib} \emph
  {et~al.}}]{Akerib:2015cja}%
  \BibitemOpen
  \bibfield  {author} {\bibinfo {author} {\bibfnamefont {D.~S.}\ \bibnamefont
  {Akerib}} \emph {et~al.} (\bibinfo {collaboration} {LZ}),\ }\href@noop {} {\
  (\bibinfo {year} {2015}{\natexlab{b}})},\ \Eprint
  {http://arxiv.org/abs/1509.02910} {arXiv:1509.02910 [physics.ins-det]}
  \BibitemShut {NoStop}%
\bibitem [{\citenamefont {Aalbers}\ \emph {et~al.}(2016)\citenamefont {Aalbers}
  \emph {et~al.}}]{Aalbers:2016jon}%
  \BibitemOpen
  \bibfield  {author} {\bibinfo {author} {\bibfnamefont {J.}~\bibnamefont
  {Aalbers}} \emph {et~al.} (\bibinfo {collaboration} {DARWIN}),\ }\href
  {\doibase 10.1088/1475-7516/2016/11/017} {\bibfield  {journal} {\bibinfo
  {journal} {JCAP}\ }\textbf {\bibinfo {volume} {1611}},\ \bibinfo {pages}
  {017} (\bibinfo {year} {2016})},\ \Eprint {http://arxiv.org/abs/1606.07001}
  {arXiv:1606.07001 [astro-ph.IM]} \BibitemShut {NoStop}%
\bibitem [{\citenamefont {Minamino}(2010)}]{Minamino:2010zz}%
  \BibitemOpen
  \bibfield  {author} {\bibinfo {author} {\bibfnamefont {A.}~\bibnamefont
  {Minamino}} (\bibinfo {collaboration} {XMASS}),\ }\bibfield  {booktitle}
  {\emph {\bibinfo {booktitle} {{Technology and instrumentation in particle
  physics. Proceedings, 1st International Conference, TIPP09, Tsukuba, Japan,
  March 12-17, 2009}}},\ }\href {\doibase 10.1016/j.nima.2010.03.032}
  {\bibfield  {journal} {\bibinfo  {journal} {Nucl. Instrum. Meth.}\ }\textbf
  {\bibinfo {volume} {A623}},\ \bibinfo {pages} {448} (\bibinfo {year}
  {2010})}\BibitemShut {NoStop}%
\bibitem [{\citenamefont {Abe}(2016)}]{Abe:2016wqw}%
  \BibitemOpen
  \bibfield  {author} {\bibinfo {author} {\bibfnamefont {K.}~\bibnamefont
  {Abe}} (\bibinfo {collaboration} {XMASS}),\ }\bibfield  {booktitle} {\emph
  {\bibinfo {booktitle} {{Proceedings, Workshop on Neutrino Physics : Session
  of CETUP* 2015 and 9th International Conference on Interconnections between
  Particle Physics and Cosmology (PPC2015): Lead/Deadwood, South Dakota, USA,
  July 6-17, 2015}}},\ }\href {\doibase 10.1063/1.4953302} {\bibfield
  {journal} {\bibinfo  {journal} {AIP Conf. Proc.}\ }\textbf {\bibinfo {volume}
  {1743}},\ \bibinfo {pages} {050001} (\bibinfo {year} {2016})}\BibitemShut
  {NoStop}%
\bibitem [{\citenamefont {Ichimura}(2016)}]{Ichimura:2015xqs}%
  \BibitemOpen
  \bibfield  {author} {\bibinfo {author} {\bibfnamefont {K.}~\bibnamefont
  {Ichimura}},\ }\bibfield  {booktitle} {\emph {\bibinfo {booktitle}
  {{Proceedings, 34th International Cosmic Ray Conference (ICRC 2015): The
  Hague, The Netherlands, July 30-August 6, 2015}}},\ }\href@noop {} {\bibfield
   {journal} {\bibinfo  {journal} {PoS}\ }\textbf {\bibinfo {volume}
  {ICRC2015}},\ \bibinfo {pages} {1223} (\bibinfo {year} {2016})}\BibitemShut
  {NoStop}%
\bibitem [{\citenamefont {Wright}(2012)}]{Wright:2012raa}%
  \BibitemOpen
  \bibfield  {author} {\bibinfo {author} {\bibfnamefont {A.}~\bibnamefont
  {Wright}} (\bibinfo {collaboration} {DarkSide}),\ }in\ \href {\doibase
  10.1142/9789814405072_0061} {\emph {\bibinfo {booktitle} {{Proceedings, 13th
  ICATPP Conference on Astroparticle, Particle, Space Physics and Detectors for
  Physics Applications (ICATPP 2011): Como, Italy, October 3-7, 2011}}}}\
  (\bibinfo {year} {2012})\ pp.\ \bibinfo {pages} {414--420}\BibitemShut
  {NoStop}%
\bibitem [{\citenamefont {Agnes}\ \emph {et~al.}(2015)\citenamefont {Agnes}
  \emph {et~al.}}]{Agnes:2014bvk}%
  \BibitemOpen
  \bibfield  {author} {\bibinfo {author} {\bibfnamefont {P.}~\bibnamefont
  {Agnes}} \emph {et~al.} (\bibinfo {collaboration} {DarkSide}),\ }\href
  {\doibase 10.1016/j.physletb.2015.03.012} {\bibfield  {journal} {\bibinfo
  {journal} {Phys. Lett.}\ }\textbf {\bibinfo {volume} {B743}},\ \bibinfo
  {pages} {456} (\bibinfo {year} {2015})},\ \Eprint
  {http://arxiv.org/abs/1410.0653} {arXiv:1410.0653 [astro-ph.CO]} \BibitemShut
  {NoStop}%
\bibitem [{\citenamefont {Agnes}\ \emph {et~al.}(2016)\citenamefont {Agnes}
  \emph {et~al.}}]{Agnes:2015ftt}%
  \BibitemOpen
  \bibfield  {author} {\bibinfo {author} {\bibfnamefont {P.}~\bibnamefont
  {Agnes}} \emph {et~al.} (\bibinfo {collaboration} {DarkSide}),\ }\href
  {\doibase 10.1103/PhysRevD.93.081101, 10.1103/PhysRevD.95.069901} {\bibfield
  {journal} {\bibinfo  {journal} {Phys. Rev.}\ }\textbf {\bibinfo {volume}
  {D93}},\ \bibinfo {pages} {081101} (\bibinfo {year} {2016})},\ \bibinfo
  {note} {[Addendum: Phys. Rev.D95,no.6,069901(2017)]},\ \Eprint
  {http://arxiv.org/abs/1510.00702} {arXiv:1510.00702 [astro-ph.CO]}
  \BibitemShut {NoStop}%
\bibitem [{\citenamefont {Davini}\ \emph {et~al.}(2016)\citenamefont {Davini},
  \citenamefont {Agnes},\ and\ \citenamefont {Franco}}]{Davini:2016vpd}%
  \BibitemOpen
  \bibfield  {author} {\bibinfo {author} {\bibfnamefont {S.}~\bibnamefont
  {Davini}}, \bibinfo {author} {\bibfnamefont {P.}~\bibnamefont {Agnes}}, \
  and\ \bibinfo {author} {\bibfnamefont {D.}~\bibnamefont {Franco}} (\bibinfo
  {collaboration} {DarkSide}),\ }\bibfield  {booktitle} {\emph {\bibinfo
  {booktitle} {{Proceedings, 14th International Conference on Topics in
  Astroparticle and Underground Physics (TAUP 2015): Torino, Italy, September
  7-11, 2015}}},\ }\href {\doibase 10.1088/1742-6596/718/4/042016} {\bibfield
  {journal} {\bibinfo  {journal} {J. Phys. Conf. Ser.}\ }\textbf {\bibinfo
  {volume} {718}},\ \bibinfo {pages} {042016} (\bibinfo {year}
  {2016})}\BibitemShut {NoStop}%
\bibitem [{\citenamefont {Amaudruz}\ \emph
  {et~al.}(2016{\natexlab{a}})\citenamefont {Amaudruz} \emph
  {et~al.}}]{Amaudruz:2014nsa}%
  \BibitemOpen
  \bibfield  {author} {\bibinfo {author} {\bibfnamefont {P.~A.}\ \bibnamefont
  {Amaudruz}} \emph {et~al.} (\bibinfo {collaboration} {DEAP}),\ }\bibfield
  {booktitle} {\emph {\bibinfo {booktitle} {{Proceedings, 37th International
  Conference on High Energy Physics (ICHEP 2014): Valencia, Spain, July 2-9,
  2014}}},\ }\href {\doibase 10.1016/j.nuclphysbps.2015.09.048} {\bibfield
  {journal} {\bibinfo  {journal} {Nucl. Part. Phys. Proc.}\ }\textbf {\bibinfo
  {volume} {273-275}},\ \bibinfo {pages} {340} (\bibinfo {year}
  {2016}{\natexlab{a}})},\ \Eprint {http://arxiv.org/abs/1410.7673}
  {arXiv:1410.7673 [physics.ins-det]} \BibitemShut {NoStop}%
\bibitem [{\citenamefont {Amaudruz}\ \emph
  {et~al.}(2016{\natexlab{b}})\citenamefont {Amaudruz} \emph
  {et~al.}}]{Amaudruz:2016qqa}%
  \BibitemOpen
  \bibfield  {author} {\bibinfo {author} {\bibfnamefont {P.~A.}\ \bibnamefont
  {Amaudruz}} \emph {et~al.} (\bibinfo {collaboration} {DEAP}),\ }\href
  {\doibase 10.1016/j.astropartphys.2016.09.002} {\bibfield  {journal}
  {\bibinfo  {journal} {Astropart. Phys.}\ }\textbf {\bibinfo {volume} {85}},\
  \bibinfo {pages} {1} (\bibinfo {year} {2016}{\natexlab{b}})},\ \Eprint
  {http://arxiv.org/abs/0904.2930} {arXiv:0904.2930 [astro-ph.IM]} \BibitemShut
  {NoStop}%
\bibitem [{\citenamefont {Fatemighomi}(2016)}]{Fatemighomi:2016ree}%
  \BibitemOpen
  \bibfield  {author} {\bibinfo {author} {\bibfnamefont {N.}~\bibnamefont
  {Fatemighomi}} (\bibinfo {collaboration} {DEAP-3600}),\ }in\ \href
  {https://inspirehep.net/record/1488107/files/arXiv:1609.07990.pdf} {\emph
  {\bibinfo {booktitle} {{35th International Symposium on Physics in Collision
  (PIC 2015) Coventry, United Kingdom, September 15-19, 2015}}}}\ (\bibinfo
  {year} {2016})\ \Eprint {http://arxiv.org/abs/1609.07990} {arXiv:1609.07990
  [physics.ins-det]} \BibitemShut {NoStop}%
\bibitem [{\citenamefont {Agnese}\ \emph {et~al.}(2014)\citenamefont {Agnese}
  \emph {et~al.}}]{Agnese:2014aze}%
  \BibitemOpen
  \bibfield  {author} {\bibinfo {author} {\bibfnamefont {R.}~\bibnamefont
  {Agnese}} \emph {et~al.} (\bibinfo {collaboration} {SuperCDMS}),\ }\href
  {\doibase 10.1103/PhysRevLett.112.241302} {\bibfield  {journal} {\bibinfo
  {journal} {Phys. Rev. Lett.}\ }\textbf {\bibinfo {volume} {112}},\ \bibinfo
  {pages} {241302} (\bibinfo {year} {2014})},\ \Eprint
  {http://arxiv.org/abs/1402.7137} {arXiv:1402.7137 [hep-ex]} \BibitemShut
  {NoStop}%
\bibitem [{\citenamefont {Agnese}\ \emph {et~al.}(2016)\citenamefont {Agnese}
  \emph {et~al.}}]{Agnese:2015nto}%
  \BibitemOpen
  \bibfield  {author} {\bibinfo {author} {\bibfnamefont {R.}~\bibnamefont
  {Agnese}} \emph {et~al.} (\bibinfo {collaboration} {SuperCDMS}),\ }\href
  {\doibase 10.1103/PhysRevLett.116.071301} {\bibfield  {journal} {\bibinfo
  {journal} {Phys. Rev. Lett.}\ }\textbf {\bibinfo {volume} {116}},\ \bibinfo
  {pages} {071301} (\bibinfo {year} {2016})},\ \Eprint
  {http://arxiv.org/abs/1509.02448} {arXiv:1509.02448 [astro-ph.CO]}
  \BibitemShut {NoStop}%
\bibitem [{\citenamefont {Kang}\ \emph {et~al.}(2013)\citenamefont {Kang} \emph
  {et~al.}}]{Kang:2013sjq}%
  \BibitemOpen
  \bibfield  {author} {\bibinfo {author} {\bibfnamefont {K.-J.}\ \bibnamefont
  {Kang}} \emph {et~al.} (\bibinfo {collaboration} {CDEX}),\ }\href {\doibase
  10.1007/s11467-013-0349-1} {\bibfield  {journal} {\bibinfo  {journal} {Front.
  Phys.(Beijing)}\ }\textbf {\bibinfo {volume} {8}},\ \bibinfo {pages} {412}
  (\bibinfo {year} {2013})},\ \Eprint {http://arxiv.org/abs/1303.0601}
  {arXiv:1303.0601 [physics.ins-det]} \BibitemShut {NoStop}%
\bibitem [{\citenamefont {Zhao}\ \emph {et~al.}(2016)\citenamefont {Zhao} \emph
  {et~al.}}]{Zhao:2016dak}%
  \BibitemOpen
  \bibfield  {author} {\bibinfo {author} {\bibfnamefont {W.}~\bibnamefont
  {Zhao}} \emph {et~al.} (\bibinfo {collaboration} {CDEX}),\ }\href {\doibase
  10.1103/PhysRevD.93.092003} {\bibfield  {journal} {\bibinfo  {journal} {Phys.
  Rev.}\ }\textbf {\bibinfo {volume} {D93}},\ \bibinfo {pages} {092003}
  (\bibinfo {year} {2016})},\ \Eprint {http://arxiv.org/abs/1601.04581}
  {arXiv:1601.04581 [hep-ex]} \BibitemShut {NoStop}%
\bibitem [{\citenamefont {Aalseth}\ \emph {et~al.}(2011)\citenamefont {Aalseth}
  \emph {et~al.}}]{Aalseth:2010vx}%
  \BibitemOpen
  \bibfield  {author} {\bibinfo {author} {\bibfnamefont {C.~E.}\ \bibnamefont
  {Aalseth}} \emph {et~al.} (\bibinfo {collaboration} {CoGeNT}),\ }\href
  {\doibase 10.1103/PhysRevLett.106.131301} {\bibfield  {journal} {\bibinfo
  {journal} {Phys. Rev. Lett.}\ }\textbf {\bibinfo {volume} {106}},\ \bibinfo
  {pages} {131301} (\bibinfo {year} {2011})},\ \Eprint
  {http://arxiv.org/abs/1002.4703} {arXiv:1002.4703 [astro-ph.CO]} \BibitemShut
  {NoStop}%
\bibitem [{\citenamefont {Barreto}\ \emph {et~al.}(2012)\citenamefont {Barreto}
  \emph {et~al.}}]{Barreto:2011zu}%
  \BibitemOpen
  \bibfield  {author} {\bibinfo {author} {\bibfnamefont {J.}~\bibnamefont
  {Barreto}} \emph {et~al.} (\bibinfo {collaboration} {DAMIC}),\ }\href
  {\doibase 10.1016/j.physletb.2012.04.006} {\bibfield  {journal} {\bibinfo
  {journal} {Phys. Lett.}\ }\textbf {\bibinfo {volume} {B711}},\ \bibinfo
  {pages} {264} (\bibinfo {year} {2012})},\ \Eprint
  {http://arxiv.org/abs/1105.5191} {arXiv:1105.5191 [astro-ph.IM]} \BibitemShut
  {NoStop}%
\bibitem [{\citenamefont {Aguilar-Arevalo}\ \emph {et~al.}(2016)\citenamefont
  {Aguilar-Arevalo} \emph {et~al.}}]{Aguilar-Arevalo:2016ndq}%
  \BibitemOpen
  \bibfield  {author} {\bibinfo {author} {\bibfnamefont {A.}~\bibnamefont
  {Aguilar-Arevalo}} \emph {et~al.} (\bibinfo {collaboration} {DAMIC}),\ }\href
  {\doibase 10.1103/PhysRevD.94.082006} {\bibfield  {journal} {\bibinfo
  {journal} {Phys. Rev.}\ }\textbf {\bibinfo {volume} {D94}},\ \bibinfo {pages}
  {082006} (\bibinfo {year} {2016})},\ \Eprint
  {http://arxiv.org/abs/1607.07410} {arXiv:1607.07410 [astro-ph.CO]}
  \BibitemShut {NoStop}%
\bibitem [{\citenamefont {Bravin}\ \emph {et~al.}(1999)\citenamefont {Bravin}
  \emph {et~al.}}]{Bravin:1999fc}%
  \BibitemOpen
  \bibfield  {author} {\bibinfo {author} {\bibfnamefont {M.}~\bibnamefont
  {Bravin}} \emph {et~al.} (\bibinfo {collaboration} {CRESST}),\ }\href
  {\doibase 10.1016/S0927-6505(99)00073-0} {\bibfield  {journal} {\bibinfo
  {journal} {Astropart. Phys.}\ }\textbf {\bibinfo {volume} {12}},\ \bibinfo
  {pages} {107} (\bibinfo {year} {1999})},\ \Eprint
  {http://arxiv.org/abs/hep-ex/9904005} {arXiv:hep-ex/9904005 [hep-ex]}
  \BibitemShut {NoStop}%
\bibitem [{\citenamefont {Angloher}\ \emph {et~al.}(2016)\citenamefont
  {Angloher} \emph {et~al.}}]{Angloher:2015ewa}%
  \BibitemOpen
  \bibfield  {author} {\bibinfo {author} {\bibfnamefont {G.}~\bibnamefont
  {Angloher}} \emph {et~al.} (\bibinfo {collaboration} {CRESST}),\ }\href
  {\doibase 10.1140/epjc/s10052-016-3877-3} {\bibfield  {journal} {\bibinfo
  {journal} {Eur. Phys. J.}\ }\textbf {\bibinfo {volume} {C76}},\ \bibinfo
  {pages} {25} (\bibinfo {year} {2016})},\ \Eprint
  {http://arxiv.org/abs/1509.01515} {arXiv:1509.01515 [astro-ph.CO]}
  \BibitemShut {NoStop}%
\bibitem [{\citenamefont {Agnese}\ \emph {et~al.}(2017)\citenamefont {Agnese}
  \emph {et~al.}}]{Agnese:2016cpb}%
  \BibitemOpen
  \bibfield  {author} {\bibinfo {author} {\bibfnamefont {R.}~\bibnamefont
  {Agnese}} \emph {et~al.} (\bibinfo {collaboration} {SuperCDMS}),\ }\href
  {\doibase 10.1103/PhysRevD.95.082002} {\bibfield  {journal} {\bibinfo
  {journal} {Phys. Rev.}\ }\textbf {\bibinfo {volume} {D95}},\ \bibinfo {pages}
  {082002} (\bibinfo {year} {2017})},\ \Eprint
  {http://arxiv.org/abs/1610.00006} {arXiv:1610.00006 [physics.ins-det]}
  \BibitemShut {NoStop}%
\bibitem [{\citenamefont {Strauss}\ \emph {et~al.}(2016)\citenamefont {Strauss}
  \emph {et~al.}}]{Strauss:2016sxp}%
  \BibitemOpen
  \bibfield  {author} {\bibinfo {author} {\bibfnamefont {R.}~\bibnamefont
  {Strauss}} \emph {et~al.},\ }\bibfield  {booktitle} {\emph {\bibinfo
  {booktitle} {{Proceedings, 14th International Conference on Topics in
  Astroparticle and Underground Physics (TAUP 2015): Torino, Italy, September
  7-11, 2015}}},\ }\href {\doibase 10.1088/1742-6596/718/4/042048} {\bibfield
  {journal} {\bibinfo  {journal} {J. Phys. Conf. Ser.}\ }\textbf {\bibinfo
  {volume} {718}},\ \bibinfo {pages} {042048} (\bibinfo {year}
  {2016})}\BibitemShut {NoStop}%
\bibitem [{\citenamefont {Alexander}\ \emph {et~al.}(2016)\citenamefont
  {Alexander} \emph {et~al.}}]{Alexander:2016aln}%
  \BibitemOpen
  \bibfield  {author} {\bibinfo {author} {\bibfnamefont {J.}~\bibnamefont
  {Alexander}} \emph {et~al.}\ }(\bibinfo {year} {2016})\ \Eprint
  {http://arxiv.org/abs/1608.08632} {arXiv:1608.08632 [hep-ph]} \BibitemShut
  {NoStop}%
\bibitem [{\citenamefont {Amole}\ \emph {et~al.}(2015)\citenamefont {Amole}
  \emph {et~al.}}]{Amole:2015lsj}%
  \BibitemOpen
  \bibfield  {author} {\bibinfo {author} {\bibfnamefont {C.}~\bibnamefont
  {Amole}} \emph {et~al.} (\bibinfo {collaboration} {PICO}),\ }\href {\doibase
  10.1103/PhysRevLett.114.231302} {\bibfield  {journal} {\bibinfo  {journal}
  {Phys. Rev. Lett.}\ }\textbf {\bibinfo {volume} {114}},\ \bibinfo {pages}
  {231302} (\bibinfo {year} {2015})},\ \Eprint
  {http://arxiv.org/abs/1503.00008} {arXiv:1503.00008 [astro-ph.CO]}
  \BibitemShut {NoStop}%
\bibitem [{\citenamefont {Amole}\ \emph
  {et~al.}(2016{\natexlab{a}})\citenamefont {Amole} \emph
  {et~al.}}]{Amole:2015pla}%
  \BibitemOpen
  \bibfield  {author} {\bibinfo {author} {\bibfnamefont {C.}~\bibnamefont
  {Amole}} \emph {et~al.} (\bibinfo {collaboration} {PICO}),\ }\href {\doibase
  10.1103/PhysRevD.93.052014} {\bibfield  {journal} {\bibinfo  {journal} {Phys.
  Rev.}\ }\textbf {\bibinfo {volume} {D93}},\ \bibinfo {pages} {052014}
  (\bibinfo {year} {2016}{\natexlab{a}})},\ \Eprint
  {http://arxiv.org/abs/1510.07754} {arXiv:1510.07754 [hep-ex]} \BibitemShut
  {NoStop}%
\bibitem [{\citenamefont {Amole}\ \emph
  {et~al.}(2016{\natexlab{b}})\citenamefont {Amole} \emph
  {et~al.}}]{Amole:2016pye}%
  \BibitemOpen
  \bibfield  {author} {\bibinfo {author} {\bibfnamefont {C.}~\bibnamefont
  {Amole}} \emph {et~al.} (\bibinfo {collaboration} {PICO}),\ }\href {\doibase
  10.1103/PhysRevD.93.061101} {\bibfield  {journal} {\bibinfo  {journal} {Phys.
  Rev.}\ }\textbf {\bibinfo {volume} {D93}},\ \bibinfo {pages} {061101}
  (\bibinfo {year} {2016}{\natexlab{b}})},\ \Eprint
  {http://arxiv.org/abs/1601.03729} {arXiv:1601.03729 [astro-ph.CO]}
  \BibitemShut {NoStop}%
\bibitem [{\citenamefont {Akerib}\ \emph {et~al.}(2016)\citenamefont {Akerib}
  \emph {et~al.}}]{Akerib:2016lao}%
  \BibitemOpen
  \bibfield  {author} {\bibinfo {author} {\bibfnamefont {D.~S.}\ \bibnamefont
  {Akerib}} \emph {et~al.} (\bibinfo {collaboration} {LUX}),\ }\href {\doibase
  10.1103/PhysRevLett.116.161302} {\bibfield  {journal} {\bibinfo  {journal}
  {Phys. Rev. Lett.}\ }\textbf {\bibinfo {volume} {116}},\ \bibinfo {pages}
  {161302} (\bibinfo {year} {2016})},\ \Eprint
  {http://arxiv.org/abs/1602.03489} {arXiv:1602.03489 [hep-ex]} \BibitemShut
  {NoStop}%
\bibitem [{\citenamefont {Fu}\ \emph {et~al.}(2017)\citenamefont {Fu} \emph
  {et~al.}}]{Fu:2016ega}%
  \BibitemOpen
  \bibfield  {author} {\bibinfo {author} {\bibfnamefont {C.}~\bibnamefont {Fu}}
  \emph {et~al.} (\bibinfo {collaboration} {PandaX-II}),\ }\href {\doibase
  10.1103/PhysRevLett.118.071301} {\bibfield  {journal} {\bibinfo  {journal}
  {Phys. Rev. Lett.}\ }\textbf {\bibinfo {volume} {118}},\ \bibinfo {pages}
  {071301} (\bibinfo {year} {2017})},\ \Eprint
  {http://arxiv.org/abs/1611.06553} {arXiv:1611.06553 [hep-ex]} \BibitemShut
  {NoStop}%
\bibitem [{\citenamefont {Mayet}\ \emph {et~al.}(2016)\citenamefont {Mayet}
  \emph {et~al.}}]{Mayet:2016zxu}%
  \BibitemOpen
  \bibfield  {author} {\bibinfo {author} {\bibfnamefont {F.}~\bibnamefont
  {Mayet}} \emph {et~al.},\ }\href {\doibase 10.1016/j.physrep.2016.02.007}
  {\bibfield  {journal} {\bibinfo  {journal} {Phys. Rept.}\ }\textbf {\bibinfo
  {volume} {627}},\ \bibinfo {pages} {1} (\bibinfo {year} {2016})},\ \Eprint
  {http://arxiv.org/abs/1602.03781} {arXiv:1602.03781 [astro-ph.CO]}
  \BibitemShut {NoStop}%
\end{thebibliography}%

\end{document}